\documentclass[10pt,letterpaper,conference]{ieeeconf}

\usepackage{calrsfs,times}
\usepackage{graphicx,psfrag,enumerate}
\usepackage{amssymb}
\usepackage{amsmath}
\usepackage{calrsfs}
\usepackage{color, epsfig, epstopdf, wrapfig}
\usepackage{paralist}
\usepackage{float,cuted}

\newtheorem{proposition}{Proposition}

\def\tr{\mathop{\rm tr}\nolimits} 
\usepackage{amsmath}
\usepackage{accents}
\newlength{\dhatheight}

\usepackage{color}
\usepackage[normalem]{ulem}
\usepackage{soul} 
\soulregister\cite7
\soulregister\ref7
\soulregister\pageref7

\IEEEoverridecommandlockouts

\begin{document}

\title{\LARGE \bf Wiener Filtering for Passive Linear Quantum
  Systems\thanks{This work was 
    supported by the Australian Research Council and the ARC Centre for Quantum
    Computation and Communication Technology.}} 

\author{V.~Ugrinovskii\thanks{V. Ugrinovskii is with the School of
    Engineering and Information Technology, University of New South Wales
    at the Australian Defence Force Academy, Canberra, ACT 2600,
    Australia, {\tt\small v.ougrinovski@adfa.edu.au}} 
  \and 
       M.~R.~James\thanks{M.~R.~James are with the ARC Centre for Quantum
    Computation and Communication Technology, Research School of Engineering,
    The Australian national University, Canberra, ACT 2601, Australia,
  {\tt\small matthew.james@anu.edu.au.}}} 

\maketitle
         
\begin{abstract}
This paper considers a version of the Wiener filtering problem for 
equalization of passive quantum linear quantum systems. We demonstrate that
taking into consideration the quantum nature of the signals
involved leads to features typically not encountered in classical
equalization problems. Most significantly, 
finding a mean-square optimal quantum equalizing filter amounts to
solving a nonconvex constrained optimization problem. We discuss two
approaches to solving this problem, both involving a relaxation of the
constraint. In both cases, unlike classical 
equalization, there is a threshold on the variance of the noise below
which an improvement of the mean-square error cannot be guaranteed.
\end{abstract}

\section{Introduction}



The task of transferring quantum information 
differs significantly from its classical
(non-quantum) counterpart, since the laws of quantum mechanics limit the 
accuracy of information transfer through quantum channels. Specifically,
the signal-to-noise ratio of possible quantum measurements on the
transmission line is limited~\cite{GG-2002}, reflecting the well known fact
that a quantum state cannot be cloned at the remote location.  
This motivates a great interest in developing systematic methodologies for
the design of optimally performing quantum communication systems. 

In the classical communication theory, optimization plays an instrumental
role in balancing various trade-offs in the design of 
classical communication systems. The most celebrated example of using
optimization in signal processing are due to N.~Wiener~\cite{Wiener-1949}
who developed a general method for reducing the effects of noise and
channel distortion through minimization of the mean square error (MSE) between
the signal and its estimate over a class of linear filters. 
This paper highlights conceptual challenges that arise when the Wiener
optimization paradigm is applied in the derivation of coherent quantum
filters, i.e., filters which themselves are quantum systems. 
To be concrete, we
restrict attention to one type of the coherent 
filtering problem concerned with equalizing distortions of quantum signals
transmitted via a quantum communication channel. Owing to the analogy with
classical channel equalization, we call this problem the \emph{quantum
  equalization problem}. The paper shows that the requirement for the filter
to be physically realizable translates into additional 
constraints which render the problem of optimizing the mean square of the 
equalization error nonconvex. 

The paper is centered around the so-called
\emph{passive} quantum equalizers. Mathematically, dynamics of a passive
quantum system in the Heisenberg picture are described by complex quantum
stochastic differential equations expressed in terms of annihilation
operators only~\cite{JG-2010}. Such systems are simple to implement
experimentally by cascading 
conventional quantum optics components such as beam splitters and optical
cavities~\cite{Nurdin-2010}. Furthermore, in a general quantum system,
passivity ensures that the system dissipates energy in the input. A
striking observation that emerges from our analysis is that passivity appears
to be a rather restrictive property in the context of equalization, in that an
optimal passive coherent  
equalizer is not always able to improve the MSE. It 
turns out that the achievable improvement  depends
on the variance of the quantum noise in the filter input signal. We
give examples which reveal a threshold on this variance 
above which the optimal passive coherent equalizer delivers an improved
MSE.

The paper is organized as follows. In the next section we present the
basics of passive linear quantum systems.  The quantum passive
equalization problem is posed in
Section~\ref{sec:equal-probl-annih}. A relaxation of the problem is
proposed in Section~\ref{sec:constr-relax}. Next, in
Section~\ref{sec:two-appr}, the problem is particularized to 
demonstrate the dependency between the power spectrum density of the
equalization error and the variance of the
system noise. Two examples of the quantum coherent filter design are presented
in that section, reflecting two approaches to optimization of the
equalization error, the first one is via direct optimization of the power
spectrum density, and the second one is using the Wiener-Hopf factorization
technique~\cite{Kailath-1981}. 
Finally, concluding remarks are given in Section~\ref{Conclusions}.

\paragraph*{Notation}
For an operator $a$ in a Hilbert space $\mathfrak{H}$, $a^*$ denotes the
Hermitian adjoint operator, and if $a$ is a complex number, $a^*$ is its
complex conjugate. Let $a=(a_1,\ldots,a_n)$ be a column vector comprised of $n$
operators (i.e., $a$ is an operator $\mathfrak{H}\to \mathfrak{H}^n$); then 
$a^\#=(a_1^*,\ldots,a_n^*)$, $a^T=(a_1^T~\ldots a_n^T)$ (i.e, the row of
operators), and $a^\dagger = (a^\#)^T$. The notation $\mathrm{col}(a,b)$
denotes the column vector obtained by concatenating vectors $a$ and $b$. 
For a complex matrix $A=(A_{ij})$,
$A^\#$, $A^T$, $A^\dagger$ denote, respectively, the matrix of complex
conjugates $(A_{ij}^*)$, 
the transpose matrix and the Hermitian adjoint matrix. $[\cdot,\cdot]$
denotes the commutator of two operators in  $\mathfrak{H}$. $\tr[\cdot]$
denotes the trace of a matrix. $I$ is the identity matrix. The quantum
expectation of an operator $V$ with respect to a state $\rho$, is
denoted $\langle V\rangle=\tr[\rho V]$~\cite{Parthasarathy-2012}.  


\section{Open linear passive quantum systems}\label{sec:open-linear-passive}

An open quantum annihilation-only system represents a linear system 
\begin{eqnarray}
  \label{dyn}
  \dot {\mathbf{a}}&=&A\mathbf{a}+B\mathbf{u}, \quad \mathbf{a}(t_0)=\mathbf{a}, \nonumber \\
  \mathbf{y}&=&C \mathbf{a}+D\mathbf{u};
\end{eqnarray}
where $A$, $B$, $C$, $D$ are complex $m\times m$, $m\times n$, $n\times m$,
$n\times n$ matrices, and $u$ is a 
(column) vector of $n$ quantum  input processes. The input is assumed to be
of the form 
\begin{equation}
  \label{eq:5}
  \mathbf{u}(t)=\mathbf{u}_0(t)+\mathbf{b}(t),
\end{equation}
where $\mathbf{b}$ is a (column) vector of $n$ quantum noise processes,
$\mathbf{b}=(\mathbf{b}_1,\ldots,\mathbf{b}_n)$, and $\mathbf{u}_0(t)$ is
an adapted 
process~\cite{HP-1984}. The noise processes can be represented as 
annihilation operators on an appropriate Fock space~\cite{HP-1984}, but
from the system theory viewpoint they can be treated as quantum Gaussian
white noise processes with zero mean, 
and
the covariance 
\begin{eqnarray}
  \label{eq:8}
  \left\langle
  \left[
    \begin{array}{c}
     \mathbf{b}(t) \\ \mathbf{b}^\# (t) 
    \end{array}
  \right]  \left[
    \begin{array}{cc}
     \mathbf{b}^\dagger(t') \\ \mathbf{b}^T (t') 
    \end{array}
  \right]\right\rangle=
  \left[
    \begin{array}{cc}
I+\Sigma_{\mathrm{b}}^T & \Pi_{\mathrm{b}} \\
 \Pi_{\mathrm{b}}^\dagger & \Sigma_{\mathrm{b}}
\end{array}\right]\delta(t-t'),
\end{eqnarray}
where $\Sigma_{\mathrm{b}}$, $\Pi_{\mathrm{b}}$ are complex matrices with
the properties that $\Sigma_{\mathrm{b}}=\Sigma_{\mathrm{b}}^\dagger$,
$\Pi_{\mathrm{b}}^T=\Pi_{\mathrm{b}}$. Along with their adjoint
(creation) operators $\mathbf{b}_j^*(t)$, the noise operators satisfy
canonical commutation relations $[\mathbf{b}_j(t),
\mathbf{b}_k^*(t')]=\delta_{jk}\delta(t-t')$, $[\mathbf{b}_j(t),
\mathbf{b}_k(t')]=0$. Here, $\delta_{jk}=0$ when $j\neq k$, and is the 
identity operator  when $j=k$; 
$\delta(t-t')$ is the $\delta$-function. The column vector $\mathbf{a}(t)=(\mathbf{a}_1(t),
\ldots, \mathbf{a}_m(t))$ represents the system modes and consists of  annihilation
operators on a certain Hilbert space $\mathfrak{H}$. 
A discussion 
about open linear quantum systems can be found in~\cite{JG-2010,GJN-2010,JNP-2008}. 
%
From now on, it will be assumed that
the pair $(A,B)$ is controllable. 

For a system of the form (\ref{dyn}) to correspond to quantum physical
dynamics, it must preserve the canonical
commutation relations during its
evolution~\cite{SP-2012,JNP-2008}. According to~\cite{MP-2011}, for the
system (\ref{dyn}) this is guaranteed if and only if there
exists a Hermitian complex matrix $\Theta$ such that 
\begin{eqnarray}
  \label{eq:2}
  A\Theta+\Theta A^\dagger+B B^\dagger=0, \quad  B=-\Theta C^\dagger, \quad
  D=I.
\end{eqnarray}
Without loss of generality we will assume from now on that the conditions
(\ref{eq:2}) are satisfied for the systems under consideration with
$\Theta=I$; this can always be achieved by an appropriate choice of
coordinates~\cite{MP-2011}. Furthermore, we will assume that the matrix $A$
is Hurwitz. 

From (\ref{dyn}), the output of the system can be represented as 
\begin{eqnarray}
\mathbf{y}(t) 
&=&C e^{A(t-t_0)}\mathbf{a} \nonumber \\
&& +\int_{t_0}^tg(t-\tau)\mathbf{u}_0(\tau)d\tau+\int_{t_0}^tg(t-\tau)\mathbf{b}(\tau)d\tau.
\label{conv}
\end{eqnarray}
Here we introduced the notation for the impulse response, associated with
the system~\cite{ZJ-2013}, 
\begin{equation}
g(t)=
\begin{cases}
Ce^{At}B+\delta(t)I, & t\ge 0, \\
0, & t<0.  
\end{cases}
\label{ch.impresp}
\end{equation}

Let us introduce the transfer function of the system (\ref{dyn}), 
\[
G(s)=C(sI-A)^{-1}B+I.
\]
Since $B=-C^\dagger$, the transfer function $G(s)$ is
square. This observation holds for all passive systems considered henceforth.
%
Furthermore,
if follows from the properties of the physical realizability~\cite{SP-2012}
that for the passive system (\ref{dyn}),
\begin{equation}
  \label{eq:1}
G(s)[G(-s^*)]^\dagger=I. 
\end{equation}

In the sequel we will be interested in stationary behaviours of the systems
under consideration. Since the matrix $A$ is assumed to be stable and
assuming that $\mathbf{u}_0(t)$ 
is stationary, the stationary component of the system output is obtained
from (\ref{conv}) by letting $t_0\to -\infty$: 
\begin{eqnarray}
\mathbf{y}(t)= \int_{-\infty}^{+\infty}g(t-\tau)\mathbf{u}_0(\tau)d\tau +\int_{-\infty}^{+\infty}g(t-\tau)\mathbf{b}(\tau)d\tau.
\label{conv.1}
\end{eqnarray}
Also, for convenience the upper limit of integration has been changed to
$+\infty$ since $g(t)$ is causal.

Consider the correlation function of stationary quantum operator processes
$\mathbf{x}_j(t)$, $\mathbf{x}_k(t)$ associated with the system,
\[
R_{\mathbf{x}_j,\mathbf{x}_k}(t)=\langle
  (\mathbf{x_j}(0)-\langle\mathbf{x_j}(0)\rangle)(\mathbf{x_k}^*(t)-\langle\mathbf{x_k}^*(t)\rangle)
  \rangle.
\] 
The corresponding power spectrum density is then
\begin{equation}
  \label{eq:27}
P_{\mathbf{x_j},\mathbf{x_k}}(i\omega)=\int_{-\infty}^{+\infty}e^{-i\omega t}
R_{\mathbf{x}_j,\mathbf{x}_k}(t) dt.  
\end{equation}
The Fourier transform is understood in the sense of tempered
distributions when $R_{\mathbf{x}_j,\mathbf{x}_k}$ is not integrable. Also,
consider the extension of $P_{\mathbf{x_j},\mathbf{x_k}}(i\omega)$ to the
complex plane, given by the bilateral Laplace transform of
$R_{\mathbf{x}_j,\mathbf{x}_k}$,
\begin{equation}
  \label{eq:28}
P_{\mathbf{x_j},\mathbf{x_k}}(s)=\int_{-\infty}^{+\infty}e^{-s t}
R_{\mathbf{x}_j,\mathbf{x}_k}(t) dt.  
\end{equation}
Often, $P_{\mathbf{x_j},\mathbf{x_k}}(s)$ is also referred to as the power
spectrum density function~\cite{Kailath-1981}, although in general it may not
be real. Since the matrix $A$ is Hurwitz, 
$P_{\mathbf{x_j},\mathbf{x_k}}(s)$ is well defined on $s=i\omega$ and 
$P_{\mathbf{x_j},\mathbf{x_k}}(s)|_{s=i\omega}=P_{\mathbf{x_j},\mathbf{x_k}}(i\omega)$,
where  the expression on the left-hand side refers to the power-spectrum
density defined in (\ref{eq:28}), and the expression on right-hand side is
defined in (\ref{eq:27}). It is easy to obtain that the power spectrum
density matrix of the output 
$\mathbf{y}(t)$,  
$P_{\mathbf{y},\mathbf{y}}(s)=(P_{\mathbf{y}_j,\mathbf{y}_k}(s))$
is related to the power spectrum density matrix of the
noise $\mathbf{b}$,
$P_{\mathbf{b},\mathbf{b}}(s)=(P_{\mathbf{b}_{j},\mathbf{b}_{k}}(s))$,
in the standard manner:
\begin{equation}
  P_{\mathbf{y},\mathbf{y}}(s)=
G(s)P_{\mathbf{b},\mathbf{b}}(i\omega)[G(-s^*)]^\dagger.  
  \label{PSD}  
\end{equation}

\section{Equalization problem for annihilation-only communication
  systems}\label{sec:equal-probl-annih}

In this section, a general equalization scheme for a passive communication
system is outlined. 

\begin{figure}[t]
\psfrag{Quantum}{}
\psfrag{channel}{$G(s)$}
\psfrag{equalizer}{$H(s)$}
  \centering
  \psfrag{b}{$u$}
  \psfrag{w}{$w$}
  \psfrag{what}{$y_w$}
  \psfrag{z}{$z$}
  \psfrag{bhat}{$\hat u$}
  \psfrag{zhat}{$\hat z$}
  \psfrag{y}{$y_u$}
  \includegraphics[width=0.9\columnwidth]{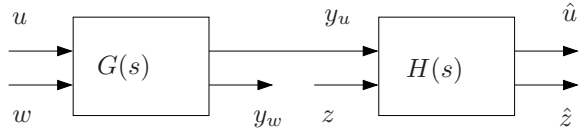}
  \caption{A general quantum communication system. The transfer function
    $G(s)$ represents the channel, and $H(s)$ represents an equalizing filter.}
  \label{fig:general}
\end{figure}

Consider a system in Fig.~\ref{fig:general} consisting of a
quantum channel and an equalizer.
The input signal $u$ plays the role of a message signal to be
transmitted through the channel, of the form 
\begin{equation}
\label{eq:18}
u(t)=u_0(t)+b(t),
\end{equation}  
and $w$ denotes the vector comprised of
additional quantum noises. It includes the noise inputs that are necessarily
present in the physically realizable system 
$G(s)$~\cite{JNP-2008,VP-2011}, as well as 
noises introduced by measurement devices. In terms of the notation
adopted in the previous section, we have
$\mathbf{u}_0=\mathrm{col}(u_0,0)$, and $\mathbf{b}=\mathrm{col}(b,w)$. 
The combined input
$\mathbf{u}=\mathrm{col}(u,w)$ drives an 
annihilation-only (passive) quantum system $G(s)$, as described in the
previous section, to produce the output
$\mathbf{y}=\mathrm{col}(y_u,y_w)$, 
although for filtering purposes, we are only interested in the output
component $y_u$ which corresponds to the input channel $u$. 

\paragraph*{The objective} 
%
%
In the classical filtering theory~\cite{Kailath-1981}, the 
equalizer is to compensate for signal distortions in the output $y_u(t)$, by
minimizing the equalization error $e(t)=\hat u(t)-u(t)$ between
classical signals $\hat u(t), u(t)$ in the mean-square
sense. The classical power spectrum density $P_{e,e}(i\omega)$ is usually
$L_2$-integrable and is related to the correlation function of the error
$e(t)$ via the inverse Fourier transform,  
\[
R_{e,e}(t)=\frac{1}{2\pi}\int_{-\infty}^{+\infty} P_{e,e}(i\omega)
e^{i\omega t}d\omega.
\]
In this case, minimizing the mean-square error covariance measure $\tr
R_{e,e}(0)$ is equivalent to minimizing $\tr P_{e,e}(i\omega)$ pointwise
in $\omega$. Alternatively, the optimal causal filter can be sought to satisfy
the Wiener-Hopf equation~\cite{Kailath-1981}, 
\begin{equation}
  \label{eq:29}
  R_{u,y_u}(t)
=\int_{0}^{+\infty}h(t-\tau)R_{y_u,y_u}(\tau)d\tau, \quad t>0; 
\end{equation}
here $h(t)$ is the unilateral inverse Laplace transform of a causal
transfer function $H(s)$. The equation (\ref{eq:29}) reflects the
projection property of classical least-square estimates, $\mathbb{E}(
e(t)y_u(\tau)^\dagger) =0$ for $-\infty< \tau<t$. The solution to
equation (\ref{eq:29}) is obtained using spectral factorization.    

Analogous to the classical mean-square equalization, we wish to
obtain a quantum system $H(s)$ whose output $\hat u$ matches the input $u$
optimally, in the sense that the equalization error $e(t)=\hat
u(t)-u(t)$ must have a minimum covariance. Owing to the physical
realizability requirement reflected in the identity~(\ref{eq:1}), quantum
channels are not guaranteed to generate $L_2$-integrable power spectrum
densities. For this reason, we will pose the problem directly as optimization of
the power spectrum density, to either minimize $\tr P_{e,e}(i\omega)$
pointwise for every $\omega$, or obtain a causal $H(s)$ by solving the
corresponding spectral factorization problem. Both approaches will be
discussed in Section~\ref{sec:two-appr}.

\paragraph*{Admissible equalizing filters}
The key distinction of the problem under consideration from classical
counterparts is that the 
system $H(s)$ must be physically realizable as a quantum system. This mandates
imposing additional requirements on the filter. 
Firstly, to ensure that the LTI filter system obtained from the optimization 
problem (\ref{eq:6}) or from spectral factorization can be made
physically realizable, it may need to be equipped with additional noise inputs
$z$ --- it was observed in~\cite{JNP-2008,VP-2011} that any LTI system can
be made physically realizable by adding noise. Without loss of generality,
we will assume that the added noise $z$ is in a Gaussian vacuum state,
i.e., the corresponding mean and covariance of $z$ are  
\begin{eqnarray}
\label{eq:12}
\langle z(t)\rangle =0, \quad 
  \left\langle
  \left[
    \begin{array}{c}
     z(t) \\ z^\# (t) 
    \end{array}
  \right]  \left[
    \begin{array}{cc}
     z^\dagger(t') \\ z^T (t') 
    \end{array}
  \right]\right\rangle=
  \left[
    \begin{array}{cc}
I & 0 \\
0 & 0
\end{array}\right]\delta(t-t').
\end{eqnarray}

Secondly, to facilitate
implementation of the resulting quantum filter~\cite{Nurdin-2010}, we
restrict attention to 
passive equalizer systems. 
In this case, the requirement for physical realizability of the filter 
leads to a formal constraint  of the form (\ref{eq:1})
on the transfer function $H(s)$:
\begin{equation}
  \label{eq:7}
  H(s)[H(-s^*)]^\dagger = I.
\end{equation}
Let us denote the set of passive physically realizable equalizers satisfying
(\ref{eq:7}) as $\mathcal{H}_{r}$. The pointwise optimization of $\tr
P_{e,e}(i\omega)$ in the class of physically realizable filters is thus a
constrained optimization problem,  
\begin{eqnarray}
  \label{eq:6}
  \label{eq:6'}
  \min_{H\in \mathcal{H}_{r}}\tr P_{e,e}(i\omega).
\end{eqnarray}

The constraint (\ref{eq:7}) precludes the direct application of
standard filtering techniques to obtain an optimal quantum Wiener
equalizer. In the next section we outline a relaxation technique which
helps to overcome this problem. 

\section{Constraint relaxation}\label{sec:constr-relax}

Let us define the partitions of the 
transfer functions $G(s)$ and $H(s)$ compatible with the 
partitions of $\mathbf{u}=\mathrm{col}(u,w)$,
$\mathbf{y}=\mathrm{col}(y_u,y_w)$, and $\mathrm{col}(y_u,z)$,
$\mathrm{col}(\hat u,\hat z)$, respectively:  
\begin{equation}
  \label{eq:98}
  G(s)=
  \left[
    \begin{array}{cc}
G_{11}(s) & G_{12}(s)\\
G_{21}(s) & G_{22}(s)\\
    \end{array}
  \right], \quad 
  H(s)=
  \left[
    \begin{array}{cc}
H_{11}(s) & H_{12}(s)\\
H_{21}(s) & H_{22}(s)
    \end{array}
  \right].
\end{equation}
With this notation, we have that
\begin{eqnarray}
\label{eq:121}
\lefteqn{P_{e,e}(s)} && \nonumber \\
&=&
(H_{11}(s)G_{11}(s)-I)(I+\Sigma_b^T)(G_{11}(-s^*)^\dagger 
H_{11}(-s^*)^\dagger -I) 
\nonumber \\
&+&
H_{11}(s)G_{12}(s)(I+\Sigma_w^T)G_{12}(-s^*)^\dagger H_{11}(-s^*)^\dagger \nonumber \\
&+&
H_{12}(s)H_{12}(-s^*)^\dagger.
\end{eqnarray}

Also, the constraint (\ref{eq:7}) is equivalent to 
\begin{eqnarray}
  \label{eq:9}
&&
H_{11}(s)H_{11}(-s^*)^\dagger+H_{12}(s)H_{12}(-s^*)^\dagger=I, \\
&&
H_{11}(s)H_{21}(-s^*)^\dagger+H_{12}(s)H_{22}(-s^*)^\dagger=0, \label{eq:10}
\\
&& 
H_{21}(s)H_{21}(-s^*)^\dagger+H_{22}(s)H_{22}(-s^*)^\dagger=I. \label{eq:11}
\end{eqnarray}
From (\ref{eq:121}), we observe that the spectral density function $
P_{e,e}(s)$ depends on the
variables $H_{11}$, $H_{12}$ only. Therefore one possible approach to solving
the equalizer design problem is to employ a two-step procedure
whose first step is to optimize the equalization error with respect
to $H_{11}(s)$, $H_{12}(s)$, subject
to the constraint (\ref{eq:9}), followed by the second step during which
the remaining transfer 
functions $H_{21}(s)$, $H_{22}(s)$ are computed to fulfill the remaining
physical realizability constraints (\ref{eq:10}), (\ref{eq:11}). 

Of course, there is no guarantee that with $H_{11}(s)$, $H_{12}(s)$ found
during the first step, the remaining transfer functions $H_{21}(s)$,
$H_{22}(s)$ exist and satisfy the conditions (\ref{eq:10}),
(\ref{eq:11}). Nevertheless, this  
approach is attractive in that it allows us to obtain tractable 
relaxations of the original quantum equalizer design problem. Indeed, using
(\ref{eq:9}), $H_{12}(s)$ can be eliminated from the expression (\ref{eq:121}): 
\begin{eqnarray}
\label{eq:13a}
P_{e,e}(s)
&=& 
(H_{11}(s)G_{11}(s)-I)(I+\Sigma_b^T) \nonumber \\
&&\times (G_{11}(-s^*)^\dagger 
H_{11}(-s^*)^\dagger -I) 
\nonumber \\
&&
+H_{11}(s)G_{12}(s)(I+\Sigma_w^T)G_{12}(-s^*)^\dagger H_{11}(-s^*)^\dagger \nonumber \\
&&
-H_{11}(s)H_{11}(-s^*)^\dagger + I.
\end{eqnarray}
It also follows from (\ref{eq:9}) that 
\begin{equation}
  \label{eq:14}
  H_{11}(i\omega)H_{11}(i\omega)^\dagger \le I \quad
  \forall\omega\in\mathbf{R}^1. 
\end{equation}
This allows us to replace the original problem of finding an optimal passive
equalizer $H(s)$ with the problem of optimizing the equalization
error in the class of causal transfer functions $H_{11}(s)$ subject to the
quadratic constraint (\ref{eq:14}). We will give a precise meaning to this
statement in the next section, where we discuss two relaxed quantum Wiener
filter problem formulations.

\section{Two approaches to quantum Wiener equalization}\label{sec:two-appr}

In this section we apply the relaxation technique discussed in the previous
section to two problems which demonstrate features of the
quantum Wiener filtering. Our aim is to
highlight new features of the problem of coherent Wiener equalization
owing to the physical realizability constraint (\ref{eq:7}), rather
than obtain a general solution to this problem. All signals in this section are
assumed to be scalar unless specified otherwise.   

\subsection{Equalization via optimization of power spectrum density: 
An optical beam splitter}

In this section, we focus on the  problem
(\ref{eq:6'}). The constraint relaxation proposed in the previous section
allows to replace this problem with the problem involving the constraint
(\ref{eq:14}). In the case of scalar signals $u$, $y_u$ and $\hat u$,
$P_{ee}(s)$ and $\Sigma_b$ are scalars, and this problems simplifies
significantly: 
\begin{eqnarray}
  \label{eq:15}
  && \min_{|H_{11}(i\omega)|\le 1} P_{e,e}(i\omega), \\
\label{eq:13b}
P_{e,e}(i\omega) &=& 
(1+\Sigma_b)|H_{11}(i\omega)G_{11}(i\omega)-1|^2 \nonumber \\
&&
+|H_{11}(i\omega)|^2G_{12}(i\omega)(I+\Sigma_w^T)G_{12}(i\omega)^\dagger \nonumber \\
&& -|H_{11}(i\omega)|^2 + 1.
\end{eqnarray}        
In (\ref{eq:15}), the minimum is taken over the set of causal
transfer functions  $H_{11}(s)$ subject to the scalar version of
the condition~(\ref{eq:14}). Obviously, we have in this case      
\begin{eqnarray}
  \label{eq:16}
   \min_{|H_{11}(i\omega)|\le 1} P_{e,e}(i\omega) &\le&
   \min_{H\in \mathcal{H}_{r}} P_{e,e}(i\omega);  
\end{eqnarray}
i.e., the problem (\ref{eq:15}) delivers a lower bound on the optimal power
spectrum density. The requirement for $H_{11}(s)$ to be causal is also
nontrivial --- while the frequency pointwise optimization is easy to
perform over complex $H_{11}$, the pointwise optimal $H_{11,\omega}$
obtained this way must admit a causal extension into the
complex plane. In general, this issue can be addressed
numerically~\cite{DGB-2015}, using the standard Matlab
software~\cite{tfest}. Therefore in the remainder  
of this section, we will be concerned with equalization of a static quantum
system for which the causality condition is satisfied automatically. This
simplified analysis aims to demonstrate that the proposed relaxation can 
lead to physically realizable equalizers 
which
are optimal in the sense of (\ref{eq:6'}).

\begin{figure}[t]
  \centering
  \psfrag{b}{$u$}
  \psfrag{w}{$w$}
  \psfrag{z}{$z$}
  \psfrag{bhat}{$\hat u$}
  \psfrag{zhat}{$\hat z$}
  \psfrag{what}{$y_w$}
  \psfrag{H}{$H(s)$}
  \psfrag{y}{$y_u$}
  \includegraphics[width=0.7\columnwidth]{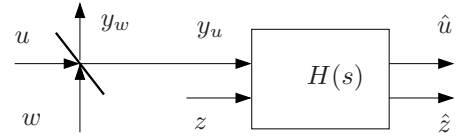}
  \caption{A beam splitter and a quantum equalizer system.}
  \label{beamsplitter}
\end{figure}

As an example of a static quantum
system consider a quantum-mechanical beam splitter, which is a
two-input two-output quantum system; see Fig.~\ref{beamsplitter}. 
In Fig.~\ref{beamsplitter}, the input $u$ represents
the signal we would like to split, and the second input $w$ is an
auxiliary noise input. The beam splitter mixes the signals $u$ and $w$, 
its outputs and inputs are related via a unitary transformation:
\begin{eqnarray}
  \label{eq:1.bs}
  \left[
    \begin{array}{c}
      y_u \\ y_w
    \end{array}
  \right]=  G \left[
    \begin{array}{c}
      u \\  w
    \end{array}
  \right], \quad G(s)=\left[
    \begin{array}{cc}\sqrt{\eta} & \sqrt{1-\eta} \\
         -\sqrt{1-\eta} & \sqrt{\eta}
    \end{array}
  \right]; \quad
\end{eqnarray}
$\eta\in(0,1)$ is a real parameter known as transmittance. That is,
$G(s)$ is static in this case, and 
\[
y_u=\sqrt{\eta}u+\sqrt{1-\eta}w.
\]
The equalization problem is to estimate the
signal $u$ from the output $y_u$ of this device using a \emph{coherent}
equalizer, i.e., a device which preserves the
canonical commutation relations.

To demonstrate the application of a quantum Wiener filter in this problem,
suppose that the input 
noise $b$ in (\ref{eq:18}) is in Gaussian vacuum state, and $\Sigma_b=0$,
$\Pi_b=0$, whereas the beamsplitter noise $w$ is in a Gaussian thermal
state, so that $\Sigma_w=\sigma_w^2>0$, $\Pi_w=0$. With these assumptions, the
expression for the objective function in (\ref{eq:13b}) becomes 
\begin{eqnarray}
  P_{e,e}(i\omega)&=&
(1-\eta)\sigma_w^2|H_{11}(i\omega)|^2 -2\sqrt{\eta}\mathrm{Re}H_{11}(i\omega)+2.
\quad~~ 
\label{e.bs}
\end{eqnarray}
The constraint condition (\ref{eq:14}) reduces in this
case to
\begin{eqnarray}
|H_{11}(i\omega)|^2\le 1. 
\label{bogo}
\end{eqnarray}

Since all coefficients in (\ref{e.bs}) are constants, the
optimal value and the optimal equalizer should also be constant. The
problem (\ref{eq:15}) is thus a regular constrained optimization problem, which 
can be solved using the Lagrange multiplier technique. 

\begin{proposition}\label{Prop1}
\begin{enumerate}[1.]
\item
If $\sigma_w^2\le \frac{\sqrt{\eta}}{(1-\eta)}$, 
then the optimal equalizer which attains minimum
in (\ref{eq:6'}) is $H(s)=I$. 
\item
On the other hand, when $\sigma_w^2> \frac{\sqrt{\eta}}{(1-\eta)}$, 
an optimal equalizer is given by
\begin{eqnarray}
  \label{eq:19}
  H_{11}(s)&=&\frac{\sqrt{\eta}}{\sigma_w^2(1-\eta)}, \quad 
  H_{12}(s)=\sqrt{1-\frac{\eta}{\sigma_w^4(1-\eta)^2}}, \nonumber \\
  H_{21}(s)&=&-H_{12}(s), \quad H_{22}(s)=H_{11}(s).
\end{eqnarray}
Such an equalizer attenuates the input $y_u$, and must include an additional
noise input $z$, to ensure that it is physical realizable.
\end{enumerate}

The corresponding expressions for the
optimal error power spectrum density are
\begin{eqnarray}
  \label{eq:23}
  \min_{H\in\mathcal{H}_r}P_{ee}=\begin{cases}
\sigma_w^2(1-\eta)-2\sqrt{\eta}+2, & \text{if $\sigma_w^2\le
  \frac{\sqrt{\eta}}{(1-\eta)}$}; \\
2-\frac{\eta}{\sigma^2(1-\eta)}, & \text{if $\sigma_w^2>
  \frac{\sqrt{\eta}}{(1-\eta)}$}.
  \end{cases} \nonumber \\
\end{eqnarray}
\end{proposition}

Comparing the power spectrum density of the error at
the input of the filter, $P_{(y_u-u),(y_u-u)}(i\omega)$, with
$P_{e,e}(i\omega)$ in (\ref{eq:23}), we observe that
$P_{(y_u-u),(y_u-u)}(i\omega)=P_{e,e}(i\omega)$ if $\sigma_w^2\le
\frac{\sqrt{\eta}}{(1-\eta)}$, and
$P_{(y_u-u),(y_u-u)}(i\omega)>P_{e,e}(i\omega)$ if $\sigma_w^2>
\frac{\sqrt{\eta}}{(1-\eta)}$. Thus, 
Proposition~\ref{Prop1} shows that the requirement for physical
realizability restricts the capacity of an optimal coherent equalizer to
respond to noise in the input signal. It is still possible to reduce
the MSE by means of a coherent equalizer, however this is
only possible  provided the covariance of the thermal noise in the
input signal is sufficiently large.  This situation differs strikingly
from the classical Wiener equalization theory.    
 

\subsection{The Wiener-Hopf technique for quantum equalization: An
  equalizer 
  for an optical cavity} 

\begin{figure}[t]
  \centering
  \psfrag{b}{\hspace{-1ex}$u$}
  \psfrag{v}{$v$}
  \psfrag{bp}{}
  \psfrag{vp}{}
  \psfrag{alpha}{$\alpha$}
  \psfrag{beta}{$\beta$}
  \psfrag{w}{\hspace{-3ex}$w$}
  \psfrag{z}{$z$}
  \psfrag{bhat}{$\hat u$}
  \psfrag{wout}{$y_w$}
  \psfrag{vout}{$v_{out}$}
  \psfrag{bout}{$u_{out}$}
  \psfrag{zhat}{$\hat z$}
  \psfrag{y}{$y_u$}
  \psfrag{H}{\hspace{-2ex}$H(s)$}
  \includegraphics[width=0.95\columnwidth]{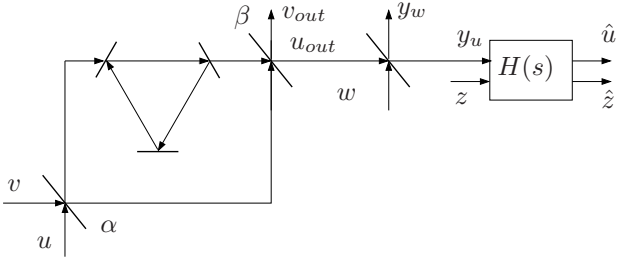}
  \caption{A cavity, beam splitters and an equalizer system.}
  \label{cavity}
\end{figure}

Let us modify the system in
Fig.~\ref{beamsplitter} to include an optical cavity and two additional
beam splitters of transmittance $\alpha$ and $\beta$; see
Fig.~\ref{cavity}. With these modification the system becomes dynamical. 
In Fig.~\ref{cavity}, $v$ denotes
 an additional thermal Gaussian noise input into the system, with zero mean
 and covariance
\[
\left\langle
  \left[
    \begin{array}{c}
     v(t) \\ v^*(t) 
    \end{array}
  \right]  \left[
    \begin{array}{cc}
     v^*(t') \\ v (t') 
    \end{array}
  \right]\right\rangle=
  \left[
    \begin{array}{cc}
1+\sigma_v^2 & 0 \\
0 & \sigma_v^2
\end{array}\right]\delta(t-t').
\]
Correspondingly, the relation between the channel output
$\mathrm{col}(y_u,y_w)$ and its input $\mathrm{col}(u,v)$ is found from the
relations  
\begin{eqnarray}
  \label{eq:1.cav}
  \left[
    \begin{array}{c}
      y_u \\ y_w
    \end{array}
  \right]&=&  \left[
    \begin{array}{cc}\sqrt{\eta}& \sqrt{1-\eta} \\
         -\sqrt{1-\eta} & \sqrt{\eta}
    \end{array}
  \right] \left[
    \begin{array}{c}
      u_{out} \\  w
    \end{array}
  \right], \nonumber \\
  \label{eq:99}
\left[
    \begin{array}{c}
      u_{out} \\ v_{out}
    \end{array}
  \right]&=& \bar G(s)\left[
    \begin{array}{c}
      u \\ v
    \end{array}
  \right], \nonumber \\
\bar G(s) &=&   \left[
    \begin{array}{cc}
\bar G_{11}(s) & \bar G_{12}(s)\\
\bar G_{21}(s) & \bar G_{22}(s)\\
    \end{array}
  \right] \nonumber \\
&=& \sqrt{\alpha\beta}\left[
    \begin{array}{cc}
      G_c-\sqrt{\alpha'\beta'}  & 
      ~~\sqrt{\alpha'}G_c+\sqrt{\beta'} \\ 
      -\sqrt{\beta'}G_c-\sqrt{\alpha'}  & 
      ~~-\sqrt{\alpha'\beta'}G_c+1  \\ 
    \end{array}
  \right];\qquad   
\end{eqnarray}
$G_c(s)$ denotes the transfer function of the optical cavity
\begin{equation}
  \label{cav.G}
  G_c(s)=\frac{s-\frac{\gamma}{2}+i\Omega}{s+\frac{\gamma}{2}+i\Omega};
\end{equation}
$\gamma$, $\Omega$ are real constants, and $
\alpha'=\frac{1-\alpha}{\alpha}$, $\beta'=\frac{1-\beta}{\beta}$. 
 Note that $G_c(s)[G_c(-s^*)]^*=I$. 

After these modifications, the power spectrum density of the equalization
error in equation~(\ref{eq:13a}) is expressed as 
\begin{eqnarray}
 P_{e,e}(s) &=& 2+ (\eta\sigma_v^2\bar G_{12}(s)[\bar
 G_{12}(-s^*)]^*+(1-\eta)\sigma_w^2)\nonumber \\
& \times& H_{11}(s)[H_{11}(-s^*)]^* \nonumber \\ 
&-& \sqrt{\eta}\left(H_{11}(s)\bar G_{11}(s)+ [H_{11}(-s^*)]^*[\bar
  G_{11}(-s^*)]^*\right). \nonumber\\
\label{e.cav}
\end{eqnarray}
The auxiliary optimization problem considered in the previous sections is
therefore to obtain a causal transfer function $H_{11}(s)$ which
optimizes (\ref{e.cav}) subject to the constraint (\ref{bogo}). 

Unlike the previous section, the system contains dynamics and the
corresponding optimal filter is expected to be dynamical. Therefore, we
cannot expect that the pointwise optimization in (\ref{eq:15}) will produce
a causal transfer function $H_{11}(s)$. In the classical case, this issue
is resolved using the Wiener-Hopf spectral factorization
method~\cite{Kailath-1981}. Therefore, here we proceed as follows. First,
we apply the Wiener-Hopf spectral factorization method~\cite{Kailath-1981}
to obtain a causal optimal $H_{11}(s)$ that minimizes $\tr
P_{e,e}(i\omega)$ for $P_{e,e}(s)$ in (\ref{e.cav}); this step does not
involve the physical realizability constraints. Next, we show that
in fact the found $H_{11}(s)$ validates the required constraint (\ref{bogo}),
provided the variance of the system noise exceeds a certain threshold. Then
we show that in this case a complete physically realizable filter transfer 
function $H(s)$ which satisfies (\ref{eq:9})--(\ref{eq:11}) can be
constructed from the found $H_{11}(s)$.   

Since $P_{e,e}(s)$ in (\ref{e.cav}) depends on $H_{11}(s)$ only, we can minimize
$\tr P_{e,e}(i\omega)$ by treating 
$P_{e,e}(s)$ as a power spectrum density of a classical system. Define  
\begin{eqnarray}
\zeta=\frac{1-\eta}{\eta}\frac{\sigma_w^2}{\alpha\beta}, \quad \rho=
\frac{\alpha'+\beta'+\frac{\zeta}{\sigma_v^2}}{2\sqrt{\alpha'\beta'}} 
\ge 1. \label{eq:140}
\end{eqnarray}
%
%
Letting $M(s)$ be the following causal transfer function, 
\begin{eqnarray}
  M(s)&=&\sqrt{2\eta\sigma_v^2\sqrt{\alpha\beta(1-\alpha)(1-\beta)}(\rho+1)}
  \nonumber \\
&& \times
  \frac{s+\frac{\gamma}{2}\sqrt{\frac{\rho-1}{\rho+1}}+i\Omega}
{s+\frac{\gamma}{2}+i\Omega},
  \label{eq:106}
\end{eqnarray}
we obtain the identity
\begin{eqnarray*}
 M(s) [M(-s^*)]^*&=&\eta\sigma_v^2\bar G_{12}(s)[\bar G_{12}(-s^*)]^* +
 (1-\eta)\sigma_w^2.
\end{eqnarray*}
Therefore,
\begin{eqnarray}
  \label{eq:22}
 P_{e,e}(s) &=& \left((M(s)H_{11}(s)-\sqrt{\eta}Q(s)\right) \nonumber \\
&& \times \left((M(-s)^*H_{11}(-s)^*-\sqrt{\eta}[Q(-s^*)]^*\right) \nonumber \\
&&- \eta Q(s)[Q(-s^*)]^* +2,
\end{eqnarray}
where 
\begin{eqnarray}
  \label{eq:26}
  Q(s)&\triangleq & \left[\frac{\bar G_{11}(-s^*)}{M(-s^*)}\right]^* \nonumber \\
&&\hspace{-1cm}=\frac{1-\sqrt{\alpha'\beta'}}{\sqrt{2\eta\sigma_v^2\sqrt{\alpha'\beta'}(\rho+1)}}
\left(1+\frac{\frac{\gamma}{2}\left(\sqrt{\frac{\rho-1}{\rho+1}}+\frac{1+\sqrt{\alpha'\beta'}}{1-\sqrt{\alpha'\beta'}}\right)}
{s+i\Omega-\frac{\gamma}{2}\sqrt{\frac{\rho-1}{\rho+1}}}\right).
\end{eqnarray}

Now consider a classical filtering problem of minimizing the
MSE between the filter output $\hat u=H(s)y_u$, where $y_u=M(s)u$,
and the signal $\bar u=\sqrt{\eta}Q(s)u$. Let $[Q(s)]_+$ denote the causal
part of $Q(s)$. 
According to the Wiener-Hopf method~\cite{Kailath-1981},
the causal solution to this problem is 
\begin{eqnarray}
  \label{eq:109}
H_{11}(s)&=&\frac{\sqrt{\eta}}{M(s)}[Q(s)]_+. 
\end{eqnarray}
This filter ensures that the error $u-\bar u$ and the filter input $y_u$
are orthogonal. Since the expression for the power spectrum density of the error
in this problem is exactly equal to the first term in (\ref{eq:22}), we
conclude that the filter (\ref{eq:109}) minimizes $P_{e,e}(i\omega)$ in the
class of causal transfer functions. 
 This yields the explicit expression for the optimal filter which is causal
 by way  of construction: 
\begin{equation}
  \label{eq:119}
  H_{11}(s)=\frac{(1-\sqrt{\alpha'\beta'})/\sqrt{\eta}}
{\sigma_v^2(\sqrt{\alpha'}+\sqrt{\beta'})^2+\zeta}
\times 
  \frac
  {s+\frac{\gamma}{2}+i\Omega}{s+\frac{\gamma}{2}\sqrt{\frac{\rho-1}{\rho+1}}+i\Omega}.
\end{equation}


\begin{proposition}\label{prop2}
Under the condition 
\begin{equation}
  \label{eq:137}
  \sigma_v^2>\frac{-\zeta(\alpha'+\beta')+\sqrt{4\zeta^2 \alpha'\beta'+\frac{(1-\sqrt{\alpha'\beta'})^2}{\eta\alpha\beta}(\alpha'-\beta')^2}}{(\alpha'-\beta')^2}
\end{equation}
the transfer function $H_{11}(s)$ in
(\ref{eq:119}) satisfies  (\ref{bogo}).
\end{proposition}

It can be shown using Proposition~\ref{prop2} that
the following constants are real under (\ref{eq:137}):
\begin{eqnarray*}
  \label{eq:128}
  \alpha_{11}&=& \frac{1-\sqrt{\alpha'\beta'}}{2\sqrt{\eta}\sigma_v^2(\rho+1)\sqrt{(1-\alpha)(1-\beta)}}, \\
  \alpha_{12}&=&\sqrt{1-\alpha_{11}^2}, \quad
  \beta_{12}=\frac{\gamma}{2}\sqrt{\frac{\rho-1}{\rho+1}-\alpha_{11}^2}. 
\label{eq:131}
\end{eqnarray*}

\begin{proposition}\label{T1}
  Suppose (\ref{eq:137}) holds. 
Then the optimal causal equalizer for
  the system under consideration in this section is given by  the following
  transfer functions
  \begin{eqnarray}
    \label{eq:133}
    H_{11}(s)&=&\alpha_{11}\frac{s+\frac{\gamma}{2}+i\Omega}{s+\frac{\gamma}{2}\sqrt{\frac{\rho-1}{\rho+1}}+i\Omega},
      \\
    H_{12}(s)&=&\frac{\alpha_{12}s+\beta_{12}+i\alpha_{12}\Omega}{s+\frac{\gamma}{2}\sqrt{\frac{\rho-1}{\rho+1}}+i\Omega},
    \label{eq:134}  \\
    H_{21}(s)&=&-\frac{\alpha_{12}s-\beta_{12}+i\alpha_{12}\Omega}{s+\frac{\gamma}{2}\sqrt{\frac{\rho-1}{\rho+1}}+i\Omega},
\label{eq:135} \\
    H_{22}(s)&=&\alpha_{11}\frac{s-\frac{\gamma}{2}+i\Omega}{s+\frac{\gamma}{2}\sqrt{\frac{\rho-1}{\rho+1}}+i\Omega}.
\label{eq:136}
  \end{eqnarray}
\end{proposition}


As we see, the condition (\ref{eq:137}) plays a critical role in the above
analysis. The expression on the right-hand side of (\ref{eq:137})
depends on $\sigma_w^2$. If
\begin{equation}
\zeta(\alpha'+\beta')> \sqrt{4\zeta^2
  \alpha'\beta'+\frac{(1-\sqrt{\alpha'\beta'})^2}{\eta\alpha\beta}(\alpha'-\beta')^2},
\label{eq:141}
\end{equation}
then this expression is negative, and (\ref{eq:137}) holds
trivially. 
%
It can be shown that if $ \sigma_w^2>
\frac{|1-\sqrt{\alpha'\beta'}|}{\sqrt{1-\eta}}$ then (\ref{eq:141}) holds,
and hence (\ref{eq:137}) is trivially satisfied. 
%
%
%
Thus we have arrived at a conclusion similar to that made in the previous
section: If the variance of the thermal noise in the system is
sufficiently large, then there exists a filter which attenuates the
thermal noise component of $y_u$ while injecting a small amount
of noise through the $z$ channel. 

\section{Conclusions}\label{Conclusions}
The paper has discussed a quantum counterpart of the classical Wiener
filtering problem for equalization of quantum systems. 
The requirement to obtain a
physically realizable passive causal equalizer imposes 
nonconvex constraints on  the filter transfer
function. We have discussed one form of relaxation
of these constraints, and have shown, via examples, that the relaxation does
not preclude finding a physically realizable coherent filter able to
reduce the signal distortion caused by the noisy quantum channel.



\bibliography{Val,irpnew}
\bibliographystyle{plain}

\end{document}